# Software and System Modeling Based on a Unified Formal Semantics

Manfred Broy, Franz Huber, Barbara Paech,
Bernhard Rumpe, and Katharina Spies

Fakultät für Informatik, Technische Universität München
{broy,huberf,paech,rumpe,spiesk}@in.tum.de

**Abstract.** Modeling and documentation are two essential ingredients for the engineering discipline of software development. During the last twenty years a wide variety of description and modeling techniques as well as document formats has been proposed. However, often these are not integrated into a coherent methodology with well-defined dependencies between the models and documentations. This hampers focused software development as well as the provision of powerful tool-support. In this paper we present the main issues and outline solutions in the direction of a unified, formal basis for software and system modeling.

## 1 Introduction

Computer technology for commercial applications has evolved rapidly from mainframes through personal computers to distributed systems. Software engineering could not keep pace with the resulting demand for powerful application development methods. This is exemplified by an ever growing number of software projects running behind schedule, delivering faulty software, not meeting users' needs, or even failing completely. There is a number of reasons for that, ranging from inadequate project management, over communication problems between domain experts and software developers to poorly documented and designed software. A recent inquiry on industrial software developers [DHP+98] has shown that despite the great variety of CASE-tools, development methods, and modeling techniques, software development still largely produces informal, incomplete and inconsistent requirements and design descriptions and poorly documented code. Modeling techniques are used selectively, but not integrated with each other or the coding. The large variety of proprietary modeling techniques and tools makes it difficult to choose an adequate selection for a project. As exemplified by the newly evolving standard *Unified Modeling Language* [BRJ97], the techniques provide a rich collection of complex notations without the corresponding semantic foundation. Since only static models are linked to code, behavioural models can only serve as illustrations not worthwhile the big effort of building the model.

This situation will only change if modeling techniques come with a set of development steps and tools for incremental model development, consistency checks, reasoning support and code generation. Mathematical description techniques like Z [Wor92] or LOTOS [Tur93] provide such development steps, but their uptake by industry is hampered by their cumbersome notation, lack of tools and lack of integration to established specification and assurance techniques [CGR93]. Recently, a number of approaches for the combination of mathematical and graphical modeling techniques has evolved (e.g. [Huß97,BHH$^+$97]) proving the viability of the integration of selected modeling techniques and formalisms. However, the integration of mathematical and graphical modeling techniques covering the whole process of system and software development is still an open problem.

The paper describes coherently the major issues in providing such an integrating basis. Experience on this subject has been gained mainly in the projects Focus [BDD$^+$93], SysLab [BGH$^+$97b] and AutoFocus [HSS96]. The project Focus is devoted to developing a mathematical development method for distributed systems. SysLab concentrates on graphical description techniques, their formal semantics based on Focus and their methodical use, in particular for object-oriented systems. AutoFocus is building a tool aimed at the development of distributed/embedded systems allowing the combined use of mathematical and graphical description techniques and providing powerful development steps based on the formal semantics. Its main application areas are components of embedded systems. None of the projects covers the whole development process, but taken together they provide a clear picture of the road to follow.

The paper is structured as follows. In the first section we introduce Focus, the theory of stream processing functions, as the mathematical basis of our work. First, we present Focus independent of a particular application area. Then we show how to adapt it to object-oriented systems. Focus comes with a set of notations and a methodology for developing formal specifications that can only be touched on in this paper. Refinement and compositionality provide the foundation for the formal development steps. We close this section with a discussion on the enhancement of formal notations to be useful for practitioners.

We then go on to describe the *indirect* use of Focus as the common formal semantics for graphical modeling techniques used in software development. We describe a set of graphical description techniques covering the main system aspects. These modeling techniques are similar to the ones used by structured or object-oriented methods. However, they differ in detail, because they were developed with a particular focus on a common formal semantics. The aim of that section is to make explicit the most important issues in providing this semantics.

The indirect use of formal methods is very valuable to the method developer. However, it is only useful to the system developer if the modeling techniques are accompanied by powerful development steps that allow to check and enforce the formal dependencies between the models. In the third section we discuss consistency checking, model validation and transformation as the most important development steps, together with possible tool support.

The modeling techniques and development steps must be integrated into a process of system development, covering requirements definition, analysis, de-sign and implementation. In the fourth section we present a framework making explicit the different modeling areas to be covered, namely the application domain, the system usage, and the software system, as well as the interplay between different system views and their corresponding modeling techniques.

We close with an outlook on future work. Related work is dicussed along the way.

## 2 Semantic Framework

In this section we describe the formal semantics as the basis for the description techniques and methodological aspects presented later. First we sketch the mathematics of system descriptions treating object-oriented systems as a special case. Then we present refinement as a major constituent of formal system development. After a short description of the formal system development process, we close with an evaluation of the direct use of Focus, our general framework for formal handling of distributed reactive systems.

### 2.1 Mathematical Basics

Focus incorporates a general semantics basis with some variants and a variety of techniques and specification formalisms based on this semantics. Here, we only give a short and informal description of the main concepts and some simple formulas. For further details, the interested reader is referred to [BS97,BDD$^+$93] for an introduction and more formalization, and [BBSS97] for an overview of case studies. Besides Focus there are many other formal development methods and description techniques like TLA, Unity or ProCoS. For further reading and a comparison between these and many other formal methods like algebraic or temporal logic approaches in combination with an uniform example we refer to [BMS96a,BMS96b].

According to the concepts of Focus, a distributed system consists of a number of components that are partially connected with each other or with the environment via one-way communication channels. Because our model is based on a discrete global time and on channels comparable with unbounded FIFO-buffers, the communication is time-synchronous and message-asynchronous. With the behaviour of each component and the topology of the network – the connection of components via the communication channels – the system is completely described: The behaviour of a system can be deduced from the behaviour of its constituents because the formal basis of Focus allows *modular* systems specification by *compositional* semantics.

**Timed Streams**

The basic data structure needed for the definition of component behaviour are *timed streams*. Assuming global and discrete time we model time flow by a special time signal $\sqrt{}$ (pronounced *tick*), indicating the end of a time interval. A timed stream is a sequence of $\sqrt{}$ and messages that contains an infinite number of time ticks. Apart from the time ticks a stream contains a finite or infinite number of messages. Let $M$ be a set of messages that does not contain the time signal $\sqrt{}$. By $M^\omega$ we denote streams of messages and by $M^{\overline{\omega}}$ the set of infinite timed streams containing an infinite number of ticks. To illustrate the concept of a timed stream we show a simple example. The timed stream

$$a \sqrt{}\ ab \sqrt{}\ \sqrt{}\ bca \sqrt{}\ b \sqrt{} \ldots$$

contains the sequence of small letters *aabbcab*. In the first time interval $a$ is communicated, in the third interval there is no communication, and in the fourth interval first $b$ then $c$ and last $a$ is communicated.

The special time signal $\sqrt{}$ should not be understood as a message that is transmitted, but as a semantic concept to represent the global time progress. Timed streams model complete communication histories: A specific stream associated with a channel between two components contains all information about *what* message is sent *when* between these components. Semantic variants of FOCUS abstract from time into the *untimed model* or describe, in the *synchronous model*, streams in which in each time interval at most one message can be transmitted between two components.

**Component Definition**

A (system) component is an active information processing unit that communicates with its environment through a set of input and output channels. To define a component, the *interface* must be declared at first. This contains a description of its input and output channels as well as the types of messages that can be received or sent via these channels. The *behaviour* of a component is described by a relation between its input and output streams fixing the set of communication histories that are valid for this component. One way to describe this relation is to define a *stream-processing function* that maps input streams to sets of output streams. Such a function reads an input stream message by message, and - as a reaction - writes output messages onto the output channels. Stream-processing functions must fulfill semantic properties like continuity, realizability, time-guardedness, as explained in the FOCUS-literature. Additionally it is possible to use state parameters to store control states or additional data and thus ease the modeling.

Let $I$ be the set of input channels and $O$ be the set of output channels. Then by $(I, O)$ the *syntactic interface* of a component is given. With every channel in $I \cup O$ we associate a data type indicating the type of messages sent on that channel.

To describe and to design the topology and the behaviour of distributed systems and their components, FOCUS offers different graphical and diagrammatical notations. All these description formalisms are well founded in the mathematical framework described in this section. A graphical representation of a component with its syntactic interface $I = \{i_1, \ldots, i_n\}$ and $O = \{o_1, \ldots, o_m\}$, and the individual channel types $S_1, \ldots, S_n$ and $R_1, \ldots, R_m$ is shown in Figure 1.

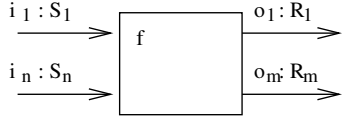

**Fig. 1.** Graphical Representation of a Component as Dataflow Node

Given a set of channels $C$ we denote the set of all channel valuations by $\boldsymbol{C}$. It is defined by:
$$\boldsymbol{C} = (C \to M^{\overline{\omega}})$$
Channel valuations are the assignments of timed streams to all channels in $C$. We assume that the streams for the channels carry only messages of the correct type specified by the interface declaration.

We describe the behaviour of a component by a stream-processing function. It defines the relation between the input streams and output streams of a component that fulfills certain conditions with respect to their timing. A stream-processing function is represented by a set-valued function on valuations of the input channels by timed streams that yields the set of histories for the output channels
$$f \ : \ \boldsymbol{I} \ \to \ \mathcal{P}(\boldsymbol{O})$$
and fulfills the timing property of *time-guardedness*. This property ensures that output histories for the first $i+1$ time intervals only depend on the input histories for the first $i$ time intervals. In other words, the processing of messages in a component takes at least one tick of time. Thus, time-guardedness axiomatizes the time flow and supports the modeling of realistic applications, since the processing of messages or the execution of computing steps always consumes time. For a precise formal definition of this property see [BS97].

### 2.2 Foundations of Object Orientation

Based on the theory given above, we have defined a set of concepts to enrich FOCUS with an object-oriented flavor. This allows us to give a formal semantics to object-oriented modeling techniques, like UML [BRJ97], as we have done in [BHH+97].

For that purpose, we have defined a *system model* in [KRB96] that characterizes our notion of object-oriented systems. Objects can be naturally viewed as components, as defined in the last section. Based on that, communication paths are defined using identifiers, where each object is associated with exactly one identifier (its identity).

In the system model, objects interact by means of *asynchronous message passing*. Asynchronous exchange of messages between the components of a system means that a message can be sent independently of the actual state of the receiver, as, e.g., in C++ or Java. To model communication between objects we use the FOCUS basic data structure of streams and stream-processing functions.

Objects encapsulate data as well as processes. *Encapsulation of a process* means that the exchange of a message does not (necessarily) imply the exchange of control: Each object is regarded as a separate process. *Encapsulation of data* means that the state of an object is not directly visible to the environment but can be accessed using explicit communication. The data part of the object defines its state. It is given in terms of typed attributes.

Objects are grouped into classes, that define the set of attributes of an object and its method interface (message interface). This allows to model the behavior of the objects of each class $c$ as stream-processing functions $f_c$ mapping input histories to sets of output histories. As usual, classes are structured by an inheritance relation $\sqsubseteq$. We thus get a natural definition of inheritance of behavior: We postulate if a class inherits from another, its possible behaviors are a subset:

$$\forall c, d : Class.\ c \sqsubseteq d \ \Rightarrow \ f_c \subseteq f_d$$

In case of method extension, this constraint is adapted to an interface refinement constraint.

Dynamic and mobile features, such as creation of new instances and change of communication structures, are also characterized as extension of Focus.

### 2.3 Refinement and Compositionality

Based on a first formal specification, the development of software and also of distributed systems goes through several development phases (or levels of abstraction). Through these phases the envisaged system or system component is described in an increasing amount of detail until a sufficiently detailed description or even an implementation of the system is obtained. The individual steps of such a process can be captured by appropriate notions of refinement. In a refinement step, parts or aspects of a system description are specified more completely or more detailed. For this purpose, FOCUS offers a powerful compositional refinement concept as well as refinement calculi. On the semantic level, refinement is modeled by logical implication. The important refinement concepts are:

**Behavioural Refinement:** The aim of this refinement is the elimination of underspecification as needed, e.g., for the specification of fault-tolerant behavior.
**Interface Refinement:** Here, the interface of a specification is refined by changing the number or types of the channels as needed, e.g., for concretization of messages or splitting communication connections between components.
**Structural Refinement:** This concept allows the development of the structure of the distributed system by refining components by networks of components.

### 2.4 A Formal System Development Process

FOCUS provides a general framework and a *methodology in the large* for formal specification and stepwise top-down development of distributed reactive systems.

The formal system development process consists of several phases of abstraction and three main development phases:

During the *Requirements Phase*, a first formalization of a given informal problem description is developed. Since the informal description is often not detailed enough, this first step of a system specification is hard to develop. It is, however, essential for the formal system development because it will be used as the basis for further development of specifications with a growing degree of accuracy in the following phases. In this step, specifications can be formalized either as trace or as functional specifications. The transition between these paradigms is formally sound and preserving correctness.

During the *Design Phase*, the essential part of the system development, the structure of a distributed system is developed by refining it up to the intended level of granularity. These formal development steps are based on the specification determined in the requirement phase and their correctness will be shown relative to the first formalization. Because the formal development of a more detailed specification possibly uncovers mistakes or unprecise properties in earlier formalizations, the top-down development is not linear but rather leads to respecifications of some parts of earlier formalizations. Only the description of system properties in a mathematical and precise manner gives a system developer the possibility to formally prove and refine system properties and descriptions. During this phase, specifications in Focus are based on the denotational semantics which models component behaviour by stream-processing functions. For the development of the specifications during the design phase, paradigms like relational and functional specifications as well as several specification styles like Assumption/Commitment[1] or equational specifications are defined. To increase its usability Focus is adapted to support various engineering oriented and practically used techniques and formalisms like tables or diagrams, see section 3. Due to the specific natures of these variants they can be used tailor-made for the solution of specific problems.

During the *Implementation Phase* the design specification is transformed into an implementation. This phase is subject of future work.

## 2.5 Further Work

Since the semantic foundation of Focus, including its development techniques, have already been explored in depth, the emphasis of further work lies in better applicability of the methodology, especially for system developers less experienced in formal methods. For that purpose, additional wide-spread description techniques, (semi-)automatic and schematic proof support have to be offered. Several techniques for describing and specifiying systems (like tables, state or system diagrams, MSC-like event traces (cf. Section 3.5), the "Assumption/Commitment" style) were successfully integrated in the methodology. With AUTOFOCUS, tool support for system development is already available, giving future case studies a new quality by offering appropriate editors, consistency checks, code generation and even simulation. Current research activities concern the enhancement of FOCUS with methodical guidelines to ease the use of the mathematical formalism, the description techniques and the development methodology for non-specialists and to support solutions for specific application fields, like the modeling of operating systems concepts in [Spi98].

---

[1] a special relational specification style where the "Assumption" formalizes the constraints about the input histories that have to be fulfilled in order to guarantee the behaviour of a component formalized by the "Commitment". For further reading see e.g. [Bro94] and [SDW95])

Case studies are an important and stimulating work for testing FOCUS in different application areas. FOCUS will be further improved, using the experience gained from the great number of case studies collected in [BFG$^+$94,BBSS97] and future studies to come.

### 2.6 On the Direct Use of Formal Description Techniques

In the last sections we have sketched a mathematical framework and the semantic basis for system specification. This allows developers to precisely describe structural and behavioural properties of the components and the composed system. As will be argued in section 3, one can hide the mathematics from developers through the use of graphical description techniques whose semantics are based on the formal framework. However not everything can be adequately expressed in diagrams. Especially behavioural properties are difficult to express. Thus for example, object-oriented specification methods typically use state transition diagrams to describe method acceptance in classes or collaboration diagrams to describe method calls between classes, but only programming language code to define the method bodies. Mathematical specification languages like FOCUS allow complete behaviour description in a much more declarative style. To be useful for practitioners, however, the notation must be simple and the specification language must be enhanced with guidelines for a systematic development of specifications. These guidelines are useful for developers formulating properties of individual systems and application areas, as well as for method developers who need to state and verify properties of the (diagrammatic) description techniques on the basis of the formal semantics.

In the following we present an example of some guidelines to write down formal specifications in FOCUS. To make formal specification techniques and methods more acceptable it is essential that the developer is in the position to concentrate on the problem and not on the correctness of the formalization. In FOCUS, equations on stream-processing functions describe the mapping of patterns of input messages to patterns of output messages. [Spi98] proposes a special strategy to formulate the required behaviour as structured text. The translation of this text into a functional equation is supported by special schemes. In the following we show such a scheme regarding a component $C$ with one input channel $In$ and one output channel $Out$, where messages of type $Integer$ flow on these channels. We require that $C$ computes the square of each input message and sends it on the output channel. For this input/output behaviour we give the following textual description:

> If the component $C$ receives a message $X \in Integer$ on input channel
> $In$, then $C$ sends as reaction the square $X^2$ as output message on output
> channel $Out$.

This structured text, which includes all information needed to specify the required behaviour, can be translated with the available schemes in the following functional equation (here $f_C$ denotes the stream-processing function modeling the behaviour of the component $C$):

$$f_C(\{In \to X\} \circ s) \;=\; \{Out \to X^2\} \circ f_C(s)$$

## 3 Description Techniques

A description technique can be best characterized as a specialized language to describe a particular view of the systems to be developed. With the Focus method, we can precisely define our notion of a system. It is an important task to define an appropriate set of description techniques which allow developers to describe properties of systems.

In the first subsection, we will describe the notion of description techniques in general, how we treat them, and what the benefits of this treatment are.

### 3.1 Description Techniques, Notations and Semantics

A description technique serves the purpose of describing particular aspects (views) of a system. There exists a variety of graphical and textual description techniques that allow to describe different aspects.

A description technique comes along with

- a concrete syntax (this is the concrete layout of all documents),
- an abstract syntax (without "syntactic sugar"),
- context conditions for wellformedness, and
- a semantics definition.

For a precisely defined description technique all four parts must be present. In case of textual notations, concrete and abstract grammars are common for the syntax, attributes on this grammar can be used for wellformedness conditions, and the semantics is usually defined as a mapping from the syntax into an appropriate semantic domain.

Similar techniques can be used for graphical notations. Each graphical notation basically defines a *language* of wellformed documents, which serves as the syntactic domain. In order to use several description techniques to describe different aspects of the same systems, semantics definitions are necessary that map the different syntactic domains onto the same semantic domain. This is the basis needed to integrate the different description techniques during development. If we map different notations onto the same semantic domain, we (meaning the *notation developer*) can compute context conditions between different notations, which ensure consistency of several views onto a system. Moreover, we can justify the correctness of translations from one notation into another one, e.g., translating Message Sequence Charts into State Machines, or generating code. Last but not least, we can justify the correctnes of refinement calculi for the given descriptions.

There are other benefits of defining a precise semantics, e.g., the developer of the semantics gains a deeper understanding of the used notations. However, usually this formal semantics definition cannot be communicated to method users, but only the (informal) interpretation of the insights [FB97]. Thus, the most important bargain of precise semantics is the possibility to automate development steps.

Since graphical techniques usually are not powerful enough to describe (or prove) every property of a system, it is often essential to translate the documents from a graphical notation into their "semantics" and use the power of the semantic formalism to specify further aspects or verify required properties. In our case, different kinds of diagrams, such as SSDs (see Section 3.2), can be translated into formulas only using concepts of Focus.

In the following, we sketch the most important notations we have dealt with. We sketch the purpose of the notation in a methodological context and the results we have achieved on that notation, such as, semantics definitions or refinement calculi that have been developed.

We emphasize that it is important to also use explanations or other informal kinds of diagrams and text during development. A good method does not only deal with formal notations but also allows the systematic treatment of informal documents.

The AutoFocus tool uses a subset of the description techniques introduced below in variations that are tailored for the development of embedded systems (see Figure 2). Graphical and textual editors are available to create and edit specifications using different views on an embedded system. Consistency between these views can be ensured, controlled by the developer any time during the development process (see Section 4.1). From sufficiently detailed specifications, executable prototypes can be generated (see Section 4.2). Implementation work on mapping graphical specifications into semantic domains, based on our theoretical work, e.g., to conduct proofs of correctness on specifications, is currently in progress (see Section 4.3).

### 3.2 System Structure Diagrams (SSD)

System Structure Diagrams as used in AutoFocus (Figure 2, upper middle) focus on the static structure of a system. They graphically exhibit the components of a system and their interconnections. They describe the glass box view of a Focus component and are therefore similar to ROOM charts [SGW94]. These diagrams focus more on the static part of a system and are not used in UML [BRJ97], where everything is assumed to be highly dynamic.

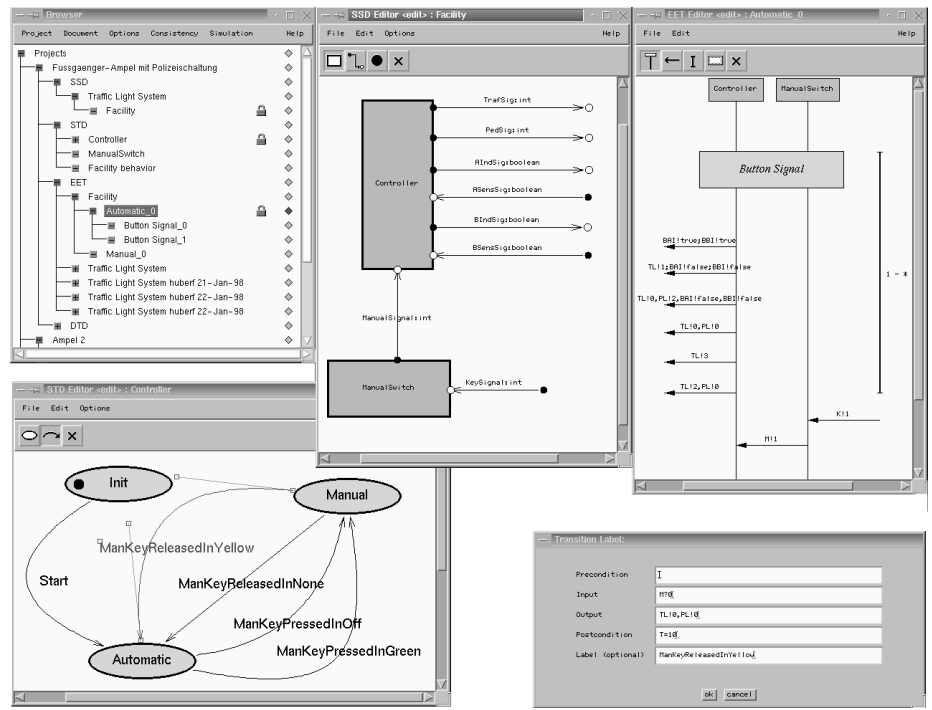

**Fig. 2.** AutoFocus Description Techniques: SSD, EET, and STD

Components may be hierachically decomposed. Therefore, for each non-elementary component an SSD can be defined, leading to a hierachy of SSD documents describing a hierachical system structure.

If a system (or system component) exhibits dynamic properties, like changing the communication structure or creating/deleting components, the SSD can be used to describe structural snapshots or the static part of the structure. In an object-oriented flavor, an SSD defines a snapshot of data and communication paths between a set of objects.

As SSDs describe the architectural part of a system, there exists a refinement calculus for architectures that allows to transform the internal structure of a component by adding new components or changing communication paths, e.g., without affecting the external behavior of the component [PR97b,PR97c].

### 3.3 Class Diagrams (CD)

Class Diagrams are the most important object-oriented notation, and are therefore part of UML [BRJ97]. They are used to describe data aspects of a system as well as possible structure layouts. In contrast to System Structure Diagrams, which focus on the "instance level", Class Diagrams focus on the "type level".

Each class may have several objects as instances, each association represents links between corresponding objects.

Class Diagrams define a large class of possible structures. To further detail these structures, different kinds of invariants are added. E.g., associations have multiplicities and additionally, it is possible to add predicates defined in our Specification Language SL (see below).

Class Diagrams are also used to define the signature of a class and their state space. The signature consists of a set of method definitions that also define the set of possible messages. The attributes define the state space.

In [BHH$^+$97] we have argued about the semantics of Class Diagrams. Although Class Diagrams are a rather well understood technique, there are still open questions how to treat aggregates.

### 3.4 Specification Languages (SL)

Not every aspect of a system can or should be described using graphic techniques. For example datatype definitions or additional constraints are best described using a textual notation. In UML, e.g., OCL has been introduced for describing a certain type of constaints. However, since OCL does not allow to define data types or auxilary functions, and based on our experiences with algebraic specification techniques [BBB$^+$85,BFG$^+$93a], we decided to define an own language for that purpose.

SL is an axiomatic specification language based on predicate logic, resembling Spectrum [BFG$^+$93a,BFG$^+$93b]. SL allows declarative definitions of properties. Particularly, SL is used for the definition of pre- and post-conditions of transitions and for the definition of state invariants not only in single objects but also between several objects in the Class Diagrams. In order to enable automatic testing of verification conditions, SL also incorporates concepts of functional programming, especially from Gofer [Jon93]. The step from high-level descriptions towards executable code is facilitated, which in turn facilitates prototyping.

With the restriction to the executable sublanguage and furthermore to the datatype definitions, an automatic translation into simulation code is possible.

We also have experimented with the higher order logic HOLCF [Reg94] as a property definition language, in particular as a front end for the theorem prover Isabelle [Pau94].

### 3.5 Message Sequence Charts (MSC) and Extended Event Traces (EET)

Message Sequence Charts and Extended Event Traces are both used to describe the flow of communication within exemplary runs of a part of a system. Constituting a high level of abstraction, MSC are well suited to capture system requirements. Moreover, MSC can be used for and generated by simulation, respectively. We have developed different flavors of this technique. One focuses on synchronous message passing between different components [BHS96,BHKS97] and its semantics is primarily a set of traces. These are called Extended Event Traces and are used in AutoFocus (Figure 2, top right).

The other variant focuses on object-oriented systems and is more similar to MSC'96 [Int96]. Both variants are compared and argued about their semantics in [BGH+97a]. For EETs a set of operators was defined to combine them sequentially, in parallel and iterated. This allows not only to define exemplary behavior, but also complete sets of behaviors.

Currently, work is in progress to map EETs into State Transition Diagrams.

### 3.6 State Transition Diagrams (STDs)

Basically State Transition Diagrams (STDs) describe the behavior of a component using the state of this component. But different abstractions and therefore flavors are possible. Thus STDs can be used early in the development (analysis) and also in the design phase, when some kind of "lifecycle" of a component is modeled. During detailed design and also prototyping, pre- and postconditions of a certain form (executable) can be used to generate code.

We have explored and developed several versions of State Transition Diagrams that allow to capture more than just one input or one output element on a transition. Usually a transition is attributed with a set of messages (sometimes restricted to one message) to be processed during the transition and a set of messages to be produced. There are timed and untimed variants, and there are variants incorporating pre- and postconditions on transitions [RK96,PR94,GKR96,GKRB96,GR95,Rum96,PR97a].

In the object-oriented flavor, State Transition Diagrams describe the lifecycle of objects. In STDs, descriptions of state and behavior are combined. STDs can be used at different levels of abstraction that allow both the specification of an object interface as well as the specification of individual methods. Refinement techniques support not only inheritance of behaviour but also stepwise refinement of abstract STDs [Rum96], resulting in an implementation.

A textual representation of State Transition Diagrams can be given using appropriate tables [Spi94,Bre97]. Hierachical variants of State Transition Diagrams are examined in [NRS96] and also used in AutoFocus (Figure 2, bottom left).

State Transition Diagrams are an extremely promising notation, as they on one hand allow to describe behavior, while on the other relate it to the state of a component. They allow to think in rather abstract terms of interaction sequences, but can also be used to describe a strategy of implementation (and therefore code generators). It is therefore worthwhile to explore more precise variants of STDs than the ones given in nowadays methods such as UML.

### 3.7 Programming Language (PL)

The ultimate description technique is the target programming language. For object-oriented systems, Java [GJS96] is a rather interesting choice for an implementation language, as it exhibits a lot of desirable properties. It is not only a language with a set of consolidated and clear concepts, it also exhibits some notion of concurrency, which allows to implement the concurrency concepts of Focus. Hence, we have had a closer look on Java, e.g., selecting a suitable sublanguage which will be the target for our code generation from STD and MSC.

To include the programming language in a proper way into the formal development process, a step has been taken in [PR97a] towards a Focus-based transitional semantics of conventional languages like Java.

### 3.8 Further Work

For some of the above described notations, we already have prototype tools—like AutoFocus—that allow to edit and manipulate documents of that notation. Several others still need consolidation, as the process of finding not only a precise semantics for given notations, but adapting the notation in such a way that it is convenient to use and allows to express the desired properties, needs to do examples.

Currently refinement calculi on Class Diagrams and State Transition Diagrams are implemented.

## 4 Methodical Ingredients

A software or system development method (see Section 5) covers a variety of different aspects. Supplying description techniques, as introduced in Section 3, is only one of these aspects, yet probably the most "visible" one. However, a development method also contains a notion of a development *process*, a model, how developers proceed during the development of a system in order to produce the results (the documents, the specifications etc.) necessary for a complete and consistent system description that fulfills the requirements and ultimately results in the desired software product.

Such a process model usually operates on different levels of granularity, ranging from a coarse view down to very detailed, even atomic operations on specification elements or documents. The former will be treated in more detail in Section 5, while the latter are covered in this section.

Methodical steps can basically be partitioned in two disjoint sets of operations on specifications, operations that modify the *contents* of specifications, thus effectively yielding a different (possibly refined) description, and operations that change the (possibly informal) *status* of specifications, for instance from a draft status to a status "validated", indicating that certain properties of the specification are fulfilled in an informal process.

In the following sections, we give a set of examples for both kinds of steps that have been treated in our work.

### 4.1 Completeness and Consistency

Generally, a system specification, just like a program that is being written, is neither complete nor consistent most of the time within a development process.

This is particularly the case in view-based systems development, which specifically aims at separating different aspects of a system description in different specification units (specification documents, for instance) that use appropriate description techniques. From a methodical point of view, allowing inconsistency and incompleteness during a development process is reasonable because enforcing them at any time restricts developers way too much in their freedom to specify systems. For instance, instead of concentrating on a certain aspect of a specification, developers, when changing parts thereof, would immediately have to update all other specification units that are possibly affected by such a change in order to maintain a consistent specification. Apart from diverting the developers' attention from their current task, this is virtually impossible in practical development, especially with respect to completeness of specifications. Note that the notion of consistency used here refers to the properties of the abstract syntax (the "meta-model") of the description techniques used to specify a system. Semantic aspects, such as consistency of behavior with certain requirements, are not treated in this context. This approach is quite similar to compilers for programming languages, which can ensure the "consistency" of a program, but not the correctness of the algorithm encoded in the program.

The AutoFocus tool, which uses a view-based approach to specify distributed systems, offers such a mechanism to test specifications for completeness and consistency. System specification is based on a subset of the description techniques introduced in Section 3, namely, system structure diagrams, datatype definitions, state transition diagrams, and extended event traces. The view specifications covered by these techniques can be developed separately to a large extent. Only at specific points in the development process, for instance, when generating a prototype from a specification (see Section 4.2), some global conditions of consistency have to be fulfilled. Consequently, the consistency mechanism available in AutoFocus is user-controlled and can be invoked at any time during development, allowing to select both an appropriate set of specifications to be checked and the (sub-)set of consistency conditions to be applied.

### 4.2 Validation of Specifications

Today in practical systems development, validation techniques, in contrast to formal verification techniques, are widely used [BCR94] to gain more confidence in specifications and implementations fulfilling their requirements. However, only verification techniques can *prove* correctness. They will be treated in the next section. Validation techniques are the focus of this section. They cover a broad range of diverse techniques, such as

- review of specifications,
- systematic specification inspection,
- (usability) test of software, or
- prototype generation and execution.

These techniques show different facets of validation. For instance, testing is usually applied to ensure that program code (the ultimate target of a development process) fulfills certain required properties. Reviews and inspections techniques, in contrast to that, are applicable in virtually any stage in the development process to ensure consistency and certain correctness aspects on an informal level. Reviews, for instance, can be held about requirements documents in the very early stages of a devlopment process as well on program code implemented by developers. Prototype generation for a system or parts thereof can be used once a specification has been developed that is sufficiently consistent and complete to validate the desired properties. Since a prototype, especially an executable prototype in the form of a program, virtually brings a system specification "into life", this kind of validation technique is relevant in communicating development results to customers. Prototyping has been successfully applied particularly in areas like graphical user interfaces (GUI).

In software engineering, the usage of graphical formalisms that describe systems from a point of view rather close to an implementation is widespread. Examples for such techniques are statecharts [HPSS87] used in the STATEMATE tool [Ilo90], or state transition diagrams as used in the AUTOFOCUS tool, both of which can basically be regarded as a kind of graphical programming language. In such cases generating executable prototypes (or as well final implementation code) is possible.

In the remainder of this section, we will take a brief look at such a prototyping environment, the AUTOFOCUS component SIMCENTER [HS97]. It is based on generating program code from a set of sufficiently detailed and consistent system specifications and on observing the behavior of that prototype program in its environment.

SIMCENTER works by generating Java program code from a specification of a distributed system, given in the AUTOFOCUS description techniques briefly outlined in Section 4.1. The generated program code, executed in SIMCENTER's runtime environment, is linked to a visualization component where the progress of the prototype execution can be monitored at the same level of description techniques as used to specify the system. An obvious prerequisite for generating such an executable prototype is that the specification is sufficiently complete and consistent in the sense outlined in Section 4.1. Nondeterminism, however, may be present in the behavioral aspects of the specification. It is currently resolved by selecting *one* possible behavior in the code generation process. This approach can be made more flexible for developers, for instance, by allowing them to select one of several nondeterministic behaviors during prototype execution.

As the primary application domain of AUTOFOCUS are embedded systems, SIMCENTER allows to monitor the interactions of such a gerated prototype with its environment. In particular, developers are able to inject stimuli into the system and observe its reactions, both from its environment interface in a black box manner and from the internal perspective, as outlined above. Additionally, black box behavior of an embedded system prototype can be optionally observed and influenced from a user-definable, application domain-oriented environment view that can be attached to SIMCENTER via a standard communication interface. This allows developers to build a very customer-oriented presentation of the be-havior of such a prototype and thus supports communication between system developers and application domain experts.

For technical details about the process and the basics of code generation in SimCenter we refer the reader to [HS97], for an AutoFocus development case study using SimCenter to validate certain correctness aspects of a specification of a simple embedded system, we refer to [HMS$^+$98].

### 4.3 Verification Techniques

In contrast to informal validation, formal techniques allow developers to mathematically prove that a system specification fulfills certain requirements. As a prerequisite, both the requirements and the specifications need to be formalized using a common mathematical basis, thus allowing formal proofs to be conducted.

Our goal is to integrate formal techniques as seamless as possible with some of the description techniques introduced in Section 3. Within the AutoFocus project two categories of verification tools are currently under consideration for an integration with graphical formalisms. First, verification systems such as PVS [ORS92], STeP [BBC$^+$96], or interactive theorem provers like *Isabelle* [Pau94] in conjunction with HOLCF [Reg94] could be used to interactively prove properties of a specification. For that purpose, graphical specifications have to be transformed into the specification laguage used in the verification system, and developers have to conduct their proofs on this notational level. Obviously, this approach is not very intuitive because it forces developers used to graphical notations to use a more or less complex mathematical formalism to conduct proofs.

Thus, the second category of tools, automated verification tools like model checkers seem to be more suitable for a seamless integration. Currently, a prototype for the integration of the $\mu$-cke model checker [Bie97] into AutoFocus is implemented. It will check whether a concrete system specification, given by a component network and the corresponding behavioral descriptions, exposes a refinement of the behavior of a given, more abstract specification.

### 4.4 Transformations

Transformations are methodical steps that effectively change a system description. Thus, each action that adds or changes specification elements results in a different system description. Whether such modifications to specifications preserve certain properties of a specification that have been established before, is not clear *a priori* and has thus again to be validated (or verified, in case of a formal development process). For that reason, it is desirable as well as feasible to have a class of methodical steps that allow developers to change specifications in a way that previously established properties will still hold after the modifications [BHS96]. Providing such property-preserving modification steps for a set of object-oriented description techniques is one of the main goals of the SysLab project. Such property-preserving transformations are defined on the level of the description techniques and provided for developers in the form of a syntactical refinement calculus that will be integrated in the toolset currently being developed within SysLab. These transformation rules are formally proven to be property-preserving by the method developers and thus enable system developers to perform transformations on specifications on the syntactical level without having to re-establish the validity of previously valid properties. Currently, such transformation calculi exist for state transition diagrams [Rum96] and for system structure diagrams [PR97b,PR97c], and are being integrated into the SysLab toolset. If developers choose not to use transformations provided by the refinement calculus, but to modify their specifications in an arbitrary way, they have to explicitly re-establish the necessary properties again.

### 4.5 Further Work

In the context of methodical development steps, tool-based active developer support is a major area of work in the near future. One aspect consists of guiding developers through the development process, offering them possible development steps that can be or must be performed in order to develop a system.

Another important aspect consists of tracing the development steps applied to specifications and their effects on other specifications. This pertains both to syntactic consistency and completeness of the specifications and to possibly invalidated semantic properties that need to be re-established after development steps.

## 5 A Model-Based Software Development Process

Up to now we have looked at formal modeling techniques, tool-support for model development and analysis based on an integrating formal basis, and a formal development process. The modeling techniques mentioned above aim at the description of the software system on various levels of granularity. In the following we show that they can naturally be complemented with a set of description techniques for the software system context and the informal problem description. We will sketch a framework for a model-based development process. This framework is made up of three main ingredients:

- the distinction between the world, the machine, and their interface [Jac95] and the explicit system models of all three of them,
- the distinction between the external view, the internal analysis view, and the (distributed) design view of each system, and
- a careful deployment of formality.

The last issue has been discussed in the preceding sections, the first two will be discussed in the following subsections. Depending on the application domain and the project context this framework needs to be instantiated. We sketch an example process for information system development at the end of this section.

### 5.1 The World, the Machine and their Interface

The distinction between *the world and the machine* is due to Jackson [Jac95]. The problem to be solved by a software system is in the world, the machine constitutes the solution we construct. Phenomena shared by the world and the machine make up the *interface*. Descriptions produced during software development must be clearly associated to one of the these three domains. This is especially difficult for requirement documents, which typically contain references to the world, namely the effects to be achieved by the software system, to the interface, namely the system services, and to the machine. In particular, it is not possible to describe the system services precisely without a clear understanding of the relevant phenomena of the world. Therefore software engineering methods - formal or pragmatic - typically start with informal descriptions of the issues in the world relevant to the software system. These are then transformed into so-called analysis models. The modeling techniques used for these models are the same as the ones used for the description of the machine. Object-oriented methods like OMT [RBP$^+$91] or OOSE [Jac92] use object models, structured methods like SSADM [DCC92] use dataflow models. This is reasonable, because the world and the machine can both be viewed as systems, thus allowing the use of the same modeling techniques. However, there are semantical differences: in object models of the software systems associations represent references directly implementable in the programming language. Associations between objects in the world represent invariant relationships which typically manifest themselves as natural phenomena (e.g., a person has a mother and a father) or as social or legal processes (e.g., a book has an author). Also, the purpose of the models of the world and the machine is quite distinct. Models of the world capture the understanding of important phenomena while models of the software system capture requirements to be realized by the software system or document the running system.

To make these distinctions explicit, we therefore distinguish three categories of models:

**Models of the world:** They model the context of the software system, e.g., a railway system or a lift to be controlled by the software system, or a production company whose engineers are supported by software systems. In particular, it is important to model the processes that the software system is involved in.

**Models of the interface:** They model the phenomena shared between the world and the machine. In particular, it is important to model the interaction between the software system and its external partners. The latter may be humans or machines.

**Models of the machine:** They model the internals of the software system, namely the internal components (e.g., objects, subsystems ) and how they render the system services.

## 5.2 The External View, the Internal View and the Design View

The world, the interface, and the machine constitute systems. They all consist of actors, communicating with each other and executing activities making use of their (data) ressources. Figure 3 collects elements of the three different systems in case of a railway control system.

|           | actors                          | data                | activities                                              |
|-----------|---------------------------------|---------------------|---------------------------------------------------------|
| **world**     | trains, passengers, conductor   | timetable, position | passengers enter and get off the train, train stops     |
| **interface** | train personnel, software system | signals             | signaling, to switch the points                         |
| **machine**   | objects, operating system processes | attributes      | assignment, method call                                 |

**Fig. 3.** The world, the interface and the machine as systems

Software development methods traditionally either focus on the activities and their data flow (structured methods) or on the actors and their communication (object-oriented methods). We claim that both views are important during system development, and that a third view has made to be explicit: the *external* view. The external view describes the services to be delivered by the system. The activities describe steps to achieve the required services. We call activities and their data the *internal analysis view* because at this level one experiments with different ways of achieving the services without regard for the actors. The actors constitute the *distributed design view*. Activities and data are encapsulated within actors such that data flow between activities has to be realized through communication. As exemplified by object-oriented designs, an actor-oriented structure allows better reusability and extensibility of designs than activity-structured designs.

Each of these views can be applied to the world, the interface, and the machine. To understand the purpose of the context of the software system, it is usually helpful to describe the services of this context. In the case of the railway control the services are the transport services offered by trains at particular locations and at particular times. In order to adequately understand the services, the activities and data of the world have to be modelled quite extensively. The actor structure of the world is frequently changed by introduction of the software system since often human labour is replaced. Furthermore, it is very often subject to a lot of political decisions.

The services of the interface are the work processes or technical processes to be supported by the software system. Jacobsen [Jac92] has coined the term *use case* for this. Very often there is a close correspondence between machine and interface services, the latter being a high-level view of the former. The internal analysis and the design view of the interface are heavily intertwined. In the interface the actors are mostly given (humans and technical systems), but there is a choice of how to distribute the activities between the machine and the external partners.

The services of the machine are determined by the design of the interface. Typically, the external view and the internal analysis view of the machine is heavily intertwined, because the services cannot be described without resorting to the data of the software system. Often, also some parts of the design view are fixed because the machine has to fit into an already existing landscape of software systems. Thus, for example, one actor may be a particular database, other actors may be given by a library of classes for a particular application domain.

### 5.3 An Example Process

The discussions above can be captured in the following proposal for the deliverables of an informations systems development process. In this short overview we do not go into detail into the dependencies between the deliverables and the possible timing of their production. The deliverables cover the external, internal, and design views for the world, the interface, and the machine. The formal system descriptions and development steps discussed in the previous sections are typically only used for the machine view. Only if the effects of the software system in the world are critical (e.g., chemical processes), formalization of the world and interface models will be worthwile.

Figure 4 lists the deliverables for developing a software system design.

| View | World | Interface | Machine |
|---|---|---|---|
| **service specification** | (textual) description of the enterprise services | use case model listing the user tasks | system services (specified in terms of their input/output and/or the data changes) |
| **data and activity analysis** | glossary, application domain processes | work processes or technical processes | data model described as ERD or CD, data changes described by STD |
| **actor and communication design** | (textual) description of the responsibility (in terms of data and activities) of the departments | (textual) description of user roles and technical system partners, allocation of data, and activities to software system | description of the component-oriented design by SSD, CD, STD, EET |

**Fig. 4.** Products of a model-based software development process

The choice of the deliverables is influenced by SSADM [DCC92], especially regarding the the machine service and analysis view. It has similarities to OOSE in the use of use cases for the external view of the interface. The use of exemplary communication flow descriptions like EETs in the machine design view is borrowed from FUSION [CAB+94].

Of course, these deliverables constitute only a framework to be instantiated for different application domains and projects. The interface models have to be quite detailed in case of human-computer interaction with a new technology [Suc95]. The world models have to be quite detailed in case of a new or critical application domain. Models of the software system should support a systematic transition to code using the development steps described in Section 4.

## 6 Conclusion

The paper has discussed the issues of using formally founded description techniques for system and software engineering. We have shown that formal methods like FOCUS provide a rich basis for textual and graphical system descriptions as well as the basic methodical steps for system development. This formal basis allows an integrated view on the wealth of description techniques found in the literature. Equally important for the system developer are the methodical elements based on the formal semantics, like consistency checks and transformations. For real-world applications, this formal development process must be embedded into a process of application domain (world) and usage (interface) understanding and description. From our experience, each of these issues is worth its own project. Our projects have demonstrated that it is possible to resolve each of these issues on its own, restricted to a particular application domain. The challenge is now to connect all of this together and to transfer it to new application domains. This can only be achieved by a widespread use of these techniques in university and industry.

## Acknowledgments

We like to thank all the people who have contributed to the work presented in this paper, especially those involved in the projects FOCUS, AUTOFOCUS, FORSOFT and SYSLAB. Furthermore, we like to thank Bernhard Schätz for his careful proof reading of the whole paper.

The authors of this paper were funded by the DFG-Sonderforschungsbereich 342, the project SYSLAB supported by DFG-Leibnitz and Siemens Nixdorf, and the Forschungsverbund FORSOFT supported by the Bayerische Forschungsstiftung.